%% file: main.tex
\documentclass[12pt]{article}

\usepackage{sbc-template}
\usepackage{graphicx,url}
\usepackage[utf8]{inputenc}
\usepackage[english]{babel}
\usepackage[T1]{fontenc}  
\usepackage{amsmath}   
\usepackage{amssymb}   
\usepackage{amsfonts}  
\usepackage{float}
\usepackage[nolist]{acronym}
\usepackage{tikz}
\usetikzlibrary{shapes, arrows, positioning, shadows, calc}
\usepackage{graphicx}
\usepackage{subfig} 
\usepackage[figurename=Figure]{caption}
\usepackage[tablename=Table]{caption}
\usepackage{tikz}
\usepackage{booktabs}
\usepackage{fontawesome}

\title{TRACE: Traceroute-based Internet Route change Analysis with Ensemble Learning}

\author{Raul Suzuki Borges\inst{1}, Rodrigo Moreira\inst{1},\\ Pedro Henrique A. Damaso de Melo\inst{1}, Larissa F. {Rodrigues Moreira}\inst{1}, \\ Flávio {de Oliveira Silva}\inst{2}}

\address{Institute of Exact and Technological Sciences -- Federal University of Viçosa
  (UFV)\\
  Rio Paranaíba -- MG -- Brazil
\nextinstitute
  Department of Informatics -- School of Engineering\\
  University of Minho (UMinho) -- Braga -- Portugal
  \email{\{raul.borges, rodrigo, larissa.f.rodrigues\}@ufv.br} \email{pedro.henrique.melo@ufv.br, flavio@di.uminho.pt}  
}

\begin{document} 
\input{acronym}

\maketitle

\begin{abstract}
Detecting Internet routing instability is a critical yet challenging task, particularly when relying solely on endpoint active measurements. This study introduces TRACE, a Machine Learning (ML) pipeline designed to identify route changes using only traceroute latency data, thereby ensuring independence from control plane information. We propose a robust feature engineering strategy that captures temporal dynamics using rolling statistics and aggregated context patterns. The architecture leverages a stacked ensemble of Gradient Boosted Decision Trees refined by a hyperparameter-optimized meta-learner. By strictly calibrating decision thresholds to address the inherent class imbalance of rare routing events, TRACE achieves a superior F1-score performance, significantly outperforming traditional baseline models and demonstrating strong effectiveness in detecting routing changes on the Internet.
\end{abstract}

\section{Introduction}\label{sec:introduction}

Enhancing network connectivity involves more than merely increasing the network capacity and speed. Instead, optimization across various layers, from the application to the routing domain, is essential to ensure stability and ultimately provide an acceptable user experience for the application~\cite{Alberti2024, Katsaros2024}. This is because different user experiences necessitate tailoring the underlying infrastructure in various forms, a concept known as \ac{NS}. \ac{NS} is defined by the \ac{3GPP} as a logical network providing specific capabilities. In a broader application context, it refers to a fully manageable segment tailored to specific verticals or business needs, as discussed in~\cite{moreira2021, Debbabi2021}. This fine management can benefit Internet routing behavior by addressing issues such as routing changes, loops, outages, Internet censorship, and even subtle attacks~\cite{Alaraj2023, Bhaskar2024}.

Advancements in programmability and virtualization have simplified \ac{NS} and, in doing so, transformed network management across many domains. The Internet is naturally collaborative and spread out, which poses challenges in sharing resources and managing them actively. Although there have been attempts to make routing programmable~\cite{moreira2021}, problems such as limited telemetry and measurement approximations still exist ~\cite{Syamkumar2022}. While \ac{AI} techniques offer great benefits~\cite{Yang2022}, measuring routing behavior remains difficult. Changes in routing can make applications unstable, leading users to think that the quality and reliability are poor. This tension leads to a concrete problem: detecting Internet routing changes without imposing heavy control-plane monitoring.

Regarding Internet behavior, specifically route changes, we propose a Taceroute-based Internet Route change Analysiswith Ensemble Learning (TRACE) approach, which is a practical avenue for identifying these events, even when hidden by rare occurrences and stochastic scenarios. The main contributions of this paper are as follows: (i)a robust feature engineering strategy that captures statistical, temporal, and aggregate context patterns from raw latency measurements; (ii)a stacked ensemble architecture that integrates LightGBM, CatBoost, and XGBoost to effectively model nonlinear routing dynamics; and (iii)a decision threshold calibration mechanism designed to maximize performance on highly imbalanced data, achieving a higher F1-score and outperforming traditional heuristics.

The remainder of this paper is organized as follows: Section~\ref{sec:related_work} reviews prior studies that closely align with our approach. In Section~\ref{sec:proposed_approach}, we present our ensemble design for identifying the routing changes. Section~\ref{sec:results_and_discussion} details our experiments, results,  In Section~\ref{sec:concluding_remarks}, we offer conclusions and suggest directions for future research that can be explored.

\section{Related Work}\label{sec:related_work}

The behavior of the Internet, particularly its routing, is intriguing, and understanding it has been the subject of study for some time~\cite{Vern1996}. This section reviews prior work on Internet routing stability, traceroute-based measurement, and path-change tracking, with emphasis on how different approaches detect and react to route dynamics. We focus on studies that analyze end-to-end path persistence, multipath and load balancing effects, reverse-path inference, and control-plane–assisted monitoring, and we summarize their main methodological traits in  Table~\ref{tab:related_work_complete} to highlight the gap addressed by TRACE.

\cite{Islam2024} addresses whether routers in the current Internet can provide consistent flow identification and path selection when endpoints rely on the Internet Protocol version six flow label together with source and destination addresses. It proposes a large scale active measurement study that extends Paris traceroute to control only network layer header fields, sending parallel traces with different flow label values from multiple worldwide vantage points to observe how load balancers actually use the Internet Protocol version six flow label in practice.

\cite{Li2022} notes that Border Gateway Protocol Multipath routing is insufficiently understood, particularly regarding how often it is deployed in the Internet and how it actually balances traffic across parallel inter domain border links. The authors mine Looking Glass server data and perform traceroute measurements from the RIPE Atlas measurement platform to infer deployments of Border Gateway Protocol Multipath and to describe the resulting multipath routing behavior and load sharing schemes.

\cite{Lin2025} examine the limited understanding of how major cloud providers configure distinct network service tiers that use either private wide area networks or the public Internet and how these choices affect user perceived latency. They design a large scale measurement study that uses geographically distributed probes and cloud regions to reveal the routing behavior of each tier and to explore how a simple performance based routing strategy could further reduce access latency.

\cite{Zakaria2020} revisits the stability and diversity of Internet routes under the widespread deployment of Multi Protocol Label Switching (MPLS), questioning whether earlier low rate measurement studies still capture current end to end path behavior. It analyzes two large traceroute data sets from PlanetLab and Ripe Atlas, using frequent probing and several path equality definitions to quantify route persistence, route prevalence and route diversity and to compare paths that traverse Multi Protocol Label Switching tunnels with paths that do not.

\cite{Schmid2025} address the challenge of characterizing transient forwarding anomalies such as routing loops and black holes that arise during the convergence process of the Border Gateway Protocol. The authors propose an inference system named TRIX that reconstructs network wide traffic flow intervals from control plane logs by simulating the decision process and modeling the Forwarding Information Base update latency.

\cite{Li2025} investigate whether reverse traceroute based on the Internet Protocol version four Record Route option can reliably reconstruct Internet paths, given that packets carrying options may traverse different router processing paths than ordinary traffic. They design a customized traceroute tool that injects various Internet Protocol options into Internet Control Message Protocol probes and run a large scale active measurement campaign from multiple worldwide vantage points to compare routing consistency between probes with options and probes without options.

\cite{Shapira2022} address the significant security challenge of detecting Border Gateway Protocol hijacking attacks that deflect traffic between endpoints and lead to man in the middle attacks. The authors introduce an unsupervised deep learning approach named AP2Vec that detects anomalies by embedding Autonomous System Numbers and Internet Protocol address prefixes into vector representations to identify functional role changes along routing paths.

\cite{Sagatov2025} investigate how to detect anomalous network behavior and distributed denial of service (DDoS) attacks using Internet Protocol performance metrics, focusing on the limitations of existing one way delay measurement tools and timestamping procedures. They propose a global monitoring architecture based on NetTestBox devices and a new owping2 utility that uses precise kernel timestamps to measure one way delay and relate sudden delay shifts and route changes to the onset of attacks.

\cite{Vermeulen2022} address the lack of visibility into reverse paths caused by routing asymmetry and the limitations of previous reverse traceroute tools regarding throughput, accuracy, and coverage. The authors present the second version of the Reverse Traceroute system which optimizes probe selection through ingress identification and traceroute atlas intersection while restricting symmetry assumptions to intradomain links to ensure high fidelity.

\cite{Giotsas2020} examine how systems that depend on traceroutes can keep their view of Internet routing accurate when probing capacity is tightly limited, since simple periodic remeasurement leaves many paths stale and wastes probes on routes that did not change. They introduce techniques that passively monitor Border Gateway Protocol updates and large public traceroute collections to infer which stored traceroutes have likely changed at the border router level, so that only those paths are selectively refreshed while the remaining traceroutes are safely reused.

\cite{Fazzion2025} address the high probing overhead and inefficiency inherent in continuous topology mapping systems that perform a complete remapping of the network path whenever a routing change is detected. The authors propose a local remapping technique named RemapRoute that reduces measurement overhead by identifying the specific Local Change Zone using a binary search mechanism and restricting the Multipath Detection Algorithm to only the affected hops.

\cite{Kirci2024} argue that the widely used Gao Rexford model of Internet routing is too coarse and cannot explain real path diversity because it represents each Autonomous System as a single node with restricted policy expressiveness. They propose a more granular routing model that incorporates internal router level topologies and richer intra domain and inter domain routing policies together with an efficient path finding algorithm that computes router level paths at Internet scale.

\cite{Wassermann2017} studies the problem of predicting Internet path changes and latency variations from traceroute measurements using supervised machine learning, motivated by the link between routing dynamics, path inflation, and performance degradation. The authors introduce NETPerfTrace, an Internet path tracking system based on decision tree ensembles and empirical feature distributions that forecasts path lifetime, number of path changes, and round trip time while outperforming a previous prediction framework.

\cite{Tian2019} addresses the challenge of accurately modeling and predicting future Border Gateway Protocol route decisions at the prefix level, since existing coarse and prior-knowledge-based models do not scale to the volume and dynamics of modern inter-domain routing data. The authors propose a data-driven deep learning framework that learns the Border Gateway Protocol decision process from large-scale control-plane traces to predict routing outcomes per prefix and assist in detecting routing dynamics and anomalies.

\begin{table}[htbp]
\caption{Comparison with Related Work.}
\label{tab:related_work_complete}
\resizebox{\textwidth}{!}{%
\begin{tabular}{lcccccc}
\hline
\textbf{Approach} & 
\textbf{\begin{tabular}[c]{@{}c@{}}Route-change \\ Detection\end{tabular}} & 
\textbf{\begin{tabular}[c]{@{}c@{}}Ensemble / \\ Advanced ML\end{tabular}} & 
\textbf{\begin{tabular}[c]{@{}c@{}}Temporal \\ Feature Eng.\end{tabular}} & 
\textbf{\begin{tabular}[c]{@{}c@{}}Imbalance \\ Handling\end{tabular}} & 
\textbf{\begin{tabular}[c]{@{}c@{}}Control-Plane \\ Independent\end{tabular}} & 
\textbf{\begin{tabular}[c]{@{}c@{}}Public Data \\ Compatible\end{tabular}} \\ \hline

\cite{Islam2024}     & \faCircleO & \faCircleO & \faCircleO & \faCircleO & \faCircleO & \faCircleO \\
\cite{Li2022}        & \faCircleO & \faCircleO & \faCircleO & \faCircleO & \faCircleO & \faCircleO \\
\cite{Lin2025}       & \faCircleO & \faCircleO & \faCircleO & \faCircleO & \faCircle  & \faCircle  \\
\cite{Zakaria2020}   & \faCircle  & \faCircleO & \faCircleO & \faCircleO & \faCircle  & \faCircle  \\
\cite{Schmid2025}    & \faCircle  & \faCircleO & \faCircleO & \faCircleO & \faCircleO & \faCircleO \\
\cite{Li2025}        & \faCircleO & \faCircleO & \faCircleO & \faCircleO & \faCircle  & \faCircleO \\
\cite{Sagatov2025}   & \faCircle  & \faCircleO & \faCircle  & \faCircleO & \faCircle  & \faCircleO \\
\cite{Vermeulen2022} & \faCircleO & \faCircleO & \faCircleO & \faCircleO & \faCircle  & \faCircle  \\
\cite{Giotsas2020}   & \faCircle  & \faCircleO & \faCircleO & \faCircleO & \faCircleO & \faCircle  \\
\cite{Fazzion2025}   & \faCircle  & \faCircleO & \faCircleO & \faCircleO & \faCircleO & \faCircleO \\
\cite{Kirci2024}     & \faCircleO & \faCircleO & \faCircleO & \faCircleO & \faCircleO & \faCircleO \\

\cite{Shapira2022}   & \faCircle  & \faCircleO & \faCircleO & \faCircleO & \faCircleO & \faCircle  \\ 
\cite{Wassermann2017}& \faCircle  & \faCircleO & \faCircle  & \faCircleO & \faCircle  & \faCircle  \\
\cite{Tian2019}      & \faCircleO & \faCircle  & \faCircleO & \faCircleO & \faCircleO & \faCircle  \\ \hline

\textbf{TRACE (Ours)} & \textbf{\faCircle} & \textbf{\faCircle} & \textbf{\faCircle} & \textbf{\faCircle} & \textbf{\faCircle} & \textbf{\faCircle} \\ \hline
\end{tabular}%
}
\end{table}

Table~\ref{tab:related_work_complete} contrasts TRACE with prior art across six key methodological dimensions, where {\scriptsize \faCircle} denotes the presence and {\scriptsize \faCircleO} the absence of a feature. The ``Route-change Detection'' column identifies works that explicitly aim to detect or predict path changes over time. ``Ensemble / Advanced ML'' distinguishes approaches that employ sophisticated learning architectures, such as stacking or deep learning, from those relying on simple heuristics or basic classifiers. ``Temporal Feature Eng.'' marks methodologies that engineer features capturing time-series dynamics, such as rolling statistics or historical deltas, rather than treating measurements as isolated snapshots. ``Imbalance Handling'' highlights studies that explicitly address class imbalance through techniques like threshold calibration, a critical aspect for rare event detection. ``Control-Plane Independent'' indicates solutions that operate solely on data-plane measurements without requiring privileged access to \ac{BGP} feeds or router configurations. Finally, ``Public Data Compatible'' denotes approaches capable of leveraging existing large-scale public datasets without necessitating bespoke active measurement infrastructure. As shown, TRACE uniquely combines these attributes to provide a robust, data-driven detection framework.

\section{Proposed Approach}\label{sec:proposed_approach}

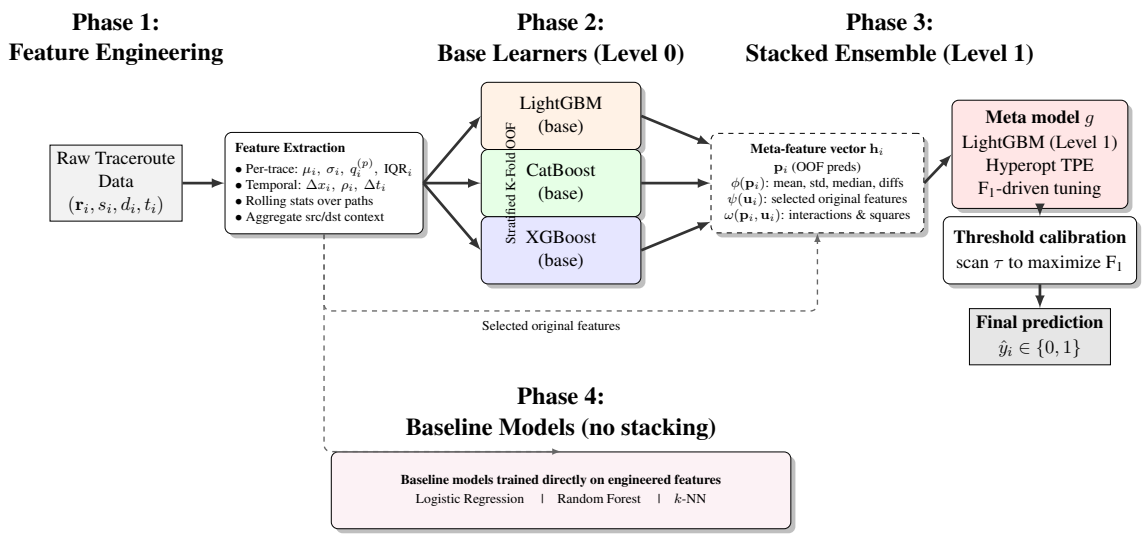
\begin{figure*}[!ht]
\centering
\resizebox{\textwidth}{!}{%
\begin{tikzpicture}[
    font=\sffamily\normalsize,
    node distance=2.5cm,
    block/.style={draw, rounded corners, align=center, fill=white,
                  drop shadow, minimum width=3.3cm, minimum height=1.4cm, inner sep=6pt},
    ds/.style={draw, align=center, minimum width=2.6cm, minimum height=1cm, fill=gray!10},
    line/.style={-latex, ultra thick, draw=black!80},
    line_skip/.style={-latex, thick, dashed, draw=black!60, rounded corners},
    label_text/.style={align=center, font=\bfseries\large}
]

\node[label_text] (ph1) at (0, 5.8) {Phase 1:\\Feature Engineering};

\node[ds, font=\small] (raw)  at (0, 2.8) {Raw Traceroute\\Data\\$(\mathbf{r}_i, s_i, d_i, t_i)$};

\node[block, align=left, font=\scriptsize] (feats) at (4.3, 2.8) {
    \textbf{Feature Extraction}\\[1mm]
    $\bullet$ Per-trace: $\mu_i,\ \sigma_i,\ q_i^{(p)},\ \text{IQR}_i$\\
    $\bullet$ Temporal: $\Delta x_i,\ \rho_i,\ \Delta t_i$\\
    $\bullet$ Rolling stats over paths\\
    $\bullet$ Aggregate src/dst context
};

\draw[line] (raw) -- (feats);

\node[label_text] (ph2) at (9.2, 5.8) {Phase 2:\\Base Learners (Level 0)};

\node[block, fill=orange!10, font=\small] (lgbm) at (9.2, 4.2) {LightGBM\\(base)};
\node[block, fill=green!10,  font=\small] (cat)  at (9.2, 2.8) {CatBoost\\(base)};
\node[block, fill=blue!10,   font=\small] (xgb)  at (9.2, 1.4) {XGBoost\\(base)};

\draw[line] (feats.east) -- (lgbm.west);
\draw[line] (feats.east) -- (cat.west);
\draw[line] (feats.east) -- (xgb.west);

\node[font=\scriptsize, rotate=90, anchor=south] at (8.4, 2.8) {Stratified K-Fold OOF};

\node[label_text] (ph3) at (16.0, 5.8) {Phase 3:\\Stacked Ensemble (Level 1)};

\node[block, dashed, minimum width=4.4cm, font=\scriptsize] (meta_vec) at (14.5, 2.8) {
    \textbf{Meta-feature vector $\mathbf{h}_i$}\\[0.5mm]
    $\mathbf{p}_i$ (OOF preds)\\
    $\phi(\mathbf{p}_i)$: mean, std, median, diffs\\
    $\psi(\mathbf{u}_i)$: selected original features\\
    $\omega(\mathbf{p}_i,\mathbf{u}_i)$: interactions \& squares
};

\draw[line] (lgbm.east) -- (meta_vec.160);
\draw[line] (cat.east)  -- (meta_vec.180);
\draw[line] (xgb.east)  -- (meta_vec.200);

\draw[line_skip] (feats.south) -- ++(0,-1.6) -| (meta_vec.south);
\node[font=\scriptsize, fill=white] at (9.0, -0.2) {Selected original features};

\node[block, fill=red!10, minimum height=1.8cm, font=\small] (meta_model) at (19.1, 3.4) {
    \textbf{Meta model $g$}\\[0.5mm]
    LightGBM (Level 1)\\
    Hyperopt TPE\\
    F\textsubscript{1}-driven tuning
};

\node[block, minimum width=3.0cm, font=\small] (thresh) at (19.1, 1.4) {
    \textbf{Threshold calibration}\\[0.5mm]
    scan $\tau$ to maximize F\textsubscript{1}
};

\node[ds, fill=black!10, font=\small] (out) at (19.1, -0.4) {
    \textbf{Final prediction}\\[0.5mm]
    $\hat{y}_i \in \{0,1\}$
};

\draw[line] (meta_vec.east) -- (meta_model.west);
\draw[line] (meta_model.south) -- (thresh.north);
\draw[line] (thresh.south) -- (out.north);

\node[label_text] (ph4) at (9.2, -2.0) {Phase 4:\\Baseline Models (no stacking)};

\node[block, fill=purple!5, minimum width=9.5cm, minimum height=1.6cm, font=\scriptsize] (baseline_panel) at (9.2, -3.6) {
    \textbf{Baseline models trained directly on engineered features}\\[0.8mm]
    Logistic Regression \quad|\quad Random Forest \quad|\quad $k$-NN
};

\draw[line_skip] (feats.south) |- (baseline_panel.north);

\end{tikzpicture}
}
\caption{Stacked ensemble architecture (Phases 1--3) and additional baseline models (Phase 4) trained directly on the engineered features for comparison.}
\label{fig:stacking_architecture}
\end{figure*}

Figure~\ref{fig:stacking_architecture} summarizes the TRACE pipeline. Starting from raw traceroute measurements, Phase~1 performs feature engineering and transforms each row into a compact vector that encodes statistical, temporal, and aggregate context. In Phase~2, these engineered features feed three gradient-boosted tree classifiers that act as base learners and produce out-of-fold (OOF) probability estimates. Phase~3 builds a meta-feature vector by combining these OOF predictions with selected original features and simple interaction terms, and trains a LightGBM meta-model on top of them. Finally, Phase~4 trains conventional baseline classifiers directly on the engineered features to isolate the benefit of stacking from the benefit of feature engineering alone.

\subsection{Data and Problem Formulation}

The traceroute dataset used in this work is drawn from the Measurement Lab (M-Lab) open repository~\cite{mlab}. For each traceroute, we have the source (\texttt{tr\_src}), the destination (\texttt{tr\_dst}), a list of round-trip times (\texttt{all\_rtts}), probe outcome counters (\texttt{total\_probes\_sent} and \texttt{total\_replies\_last\_hop}), and the binary label \texttt{route\_changed} indicating whether a route change was detected.

After preprocessing, the training set contains $N$ observations. Each observation is represented by an engineered feature vector $\mathbf{x}_i$ and a binary label $y_i \in \{0,1\}$, where $y_i = 1$ denotes a route change. The goal is to learn a scoring function
\[
f(\mathbf{x}_i) \approx \Pr(y_i = 1 \mid \mathbf{x}_i),
\]
which we later convert into binary decisions through a calibrated threshold, as detailed in the following subsections.

To evaluate the proposed solution, we use a large-scale traceroute dataset comprising a total of 28{,}521{,}656 instances. We adopt a 70\% / 30\% split in terms of number of rows, with 19{,}965{,}159 samples for training and 8{,}556{,}497 samples for testing. Each data instance corresponds to a traceroute execution characterized by the source and destination identifiers, its round-trip time (RTT) vector, and the total number of probes sent, which averages 1.44 probes per traceroute.

\begin{table}[htbp]
\caption{Dataset statistics and class distribution.}
\label{tab:dataset_stats}
\centering
\begin{tabular}{lrrr}
\hline
\textbf{Category} & \textbf{Count} & \textbf{Percentage} & \textbf{Description} \\ \hline
Total training & 19{,}965{,}159 & 100.00\% & Full training set \\
\quad \textit{Class 0 (stable)} & 19{,}595{,}589 & 98.15\% & Majority class \\
\quad \textit{Class 1 (changed)} & 369{,}570 & 1.85\% & Minority class (target) \\ \hline
Total testing & 8{,}556{,}497 & -- & Unlabeled split (30\%) \\ \hline
\textbf{Imbalance ratio} & \multicolumn{3}{c}{\textbf{1 : 53.02}} \\ \hline
\end{tabular}
\end{table}

\textbf{Class imbalance challenge.} A critical characteristic of this dataset is its severe class imbalance. As summarized in Table~\ref{tab:dataset_stats}, the target class (\texttt{route\_changed} = 1) represents only 1.85\% of the training data, whereas stable routes (\texttt{route\_changed} = 0) account for 98.15\%. This yields an imbalance ratio of approximately 1:53, which poses a significant challenge for standard classifiers that tend to bias predictions towards the majority class.

\subsection{Feature engineering (Phase 1)}

Phase~1 in Figure~\ref{fig:stacking_architecture} converts raw traceroute rows into feature vectors that capture per-trace statistics, path-level temporal dynamics, and aggregate context for sources and destinations.

\subsubsection{Per-trace statistics}

Each traceroute row contains a list of round-trip time samples, denoted by $\mathbf{r}_i = (r_{i,1}, \dots, r_{i,L_i})$. We first compress this list into stable summary statistics. The mean and variance are defined as
\begin{align}
  \mu_i &= \frac{1}{L_i} \sum_{\ell=1}^{L_i} r_{i,\ell}, \\
  \sigma_i^2 &= \frac{1}{L_i} \sum_{\ell=1}^{L_i} \bigl(r_{i,\ell} - \mu_i\bigr)^2,
\end{align}
so that $\mu_i$ reflects the typical latency of that traceroute and $\sigma_i$ quantifies its variability. In addition, we compute empirical percentiles (e.g., 25th, 50th, 75th, and 90th), the minimum and maximum latency, the length $L_i$, and the interquartile range, which together characterize different regions of the latency distribution.

Reliability features are derived from the probe counters. From the number of probes sent and the number of successful replies at the last hop, we obtain the success rate and the corresponding loss rate. These variables summarize how often the traceroute reaches the destination and how frequently probes are dropped. All these statistics, together with the original scalar attributes, form the first block of the feature vector.

\subsubsection{Path-level temporal context}

Traceroute rows that share the same source and destination form a time-ordered sequence when sorted by their measurement time. For each path $(s_i, d_i)$ we exploit this order to derive temporal features that describe how recent measurements differ from previous ones.

Consider a scalar feature $x_i$ computed from a single row, such as the mean latency or the success rate. Within a fixed path, we sort rows by increasing time and, for each row $i$ that is not the first, identify its immediate predecessor $j$ with the same source and destination. We then define the absolute and relative change as
\begin{align}
  \Delta x_i &= x_i - x_j, \\
  \rho_i &= \frac{x_i}{x_j + \varepsilon},
\end{align}
where $\varepsilon$ is a small constant to avoid division by zero. The difference $\Delta x_i$ captures the direction and magnitude of the most recent change, while the ratio $\rho_i$ measures this change relative to the previous level and is clipped to a fixed interval to limit the effect of outliers. The time interval between the two measurements is
\[
\Delta t_i = t_i - t_j,
\]
which indicates how long the path remained unobserved before the current traceroute. For the first observation of a path, these temporal features are set to zero.

\subsubsection{Rolling and aggregate context features}

In addition to local differences between consecutive rows, TRACE includes features that capture short-term trends and how typical each observation is for its source and destination.

For each path, we compute rolling means and standard deviations of selected features (such as mean latency and success rate) over windows of three and seven observations. These rolling statistics highlight whether latency or loss is gradually increasing, decreasing, or stable over recent traceroutes, without requiring the reader to follow explicit formulae.

At a coarser level, we aggregate information per source and per destination. For each source, we count how many traceroutes it originates and how many distinct destinations it reaches; the same is done per destination. We then compute empirical averages and standard deviations of key features for all rows that share a given source or destination and derive standardized deviations. For example, for a feature $x_i$ and the mean and standard deviation associated with the source of row $i$, we define
\[
z_i^{x,\text{src}} = \frac{x_i - \mu_{s_i}^{x}}{\sigma_{s_i}^{x} + \varepsilon}.
\]
Analogous quantities are computed with respect to destinations. These aggregate features indicate whether a particular measurement is unusually slow, variable, or unreliable for that source or destination, which can be a strong signal of route changes.

\subsection{Base learners and baseline models (Phases 2 and 4)}

After feature engineering, each row $i$ is represented by a vector $\mathbf{u}_i \in \mathbb{R}^{d}$, obtained by concatenating per-trace statistics, temporal context, and aggregate features. Phase~2 in Figure~\ref{fig:stacking_architecture} trains three gradient-boosted tree classifiers on this representation, implemented with LightGBM, CatBoost, and XGBoost. These models act as base learners and produce probability scores for route change.

To obtain reliable meta-features for stacking, we use stratified five-fold cross-validation. The indices $\{1,\dots,N\}$ are partitioned into five folds; for each fold and each base learner, we train on four folds and compute predictions on the held-out fold. By concatenating the predictions across all folds, we obtain three vectors of out-of-fold probabilities $p_i^{(m)}$, one per base learner, which are used as inputs to the meta-model. For the test set, we train one model per fold and average their predictions.

Phase~4 in Figure~\ref{fig:stacking_architecture} trains conventional baseline classifiers directly on the engineered feature vectors $\mathbf{u}_i$. Specifically, we consider logistic regression, Random Forest, and $k$-nearest neighbours, each tuned with a small grid over its main hyperparameters. These baselines share the same training–validation splits as the base learners and are later evaluated under the same F\textsubscript{1}-oriented threshold calibration procedure as the stacked ensemble.

\subsection{Stacked ensemble and meta-features (Phase 3)}

Phase~3 combines the strengths of the base learners through stacking. For each training row $i$, we collect the out-of-fold predictions into the vector
\[
\mathbf{p}_i = \bigl[p_i^{(1)}, p_i^{(2)}, p_i^{(3)}\bigr],
\]
which summarizes how the three base models assess the probability of a route change. From this vector, we compute a small set of summary statistics, such as the mean, standard deviation, median, and pairwise differences. These statistics, denoted by $\phi(\mathbf{p}_i)$ in Figure~\ref{fig:stacking_architecture}, measure how much the base learners agree or disagree for each example.

In addition, we select a subset of the original features, denoted by $\psi(\mathbf{u}_i)$, that were empirically found to be particularly informative, including mean latency, variability, high percentiles, success rate, and a few temporal and rolling features. Finally, we construct simple interaction features, denoted by $\omega(\mathbf{p}_i,\mathbf{u}_i)$, that capture nonlinear relationships between predictions and original variables, for example products of base predictions with the success rate and squared prediction terms.

The complete meta-feature vector for row $i$ is then
\[
\mathbf{h}_i = \bigl[\, \mathbf{p}_i,\, \phi(\mathbf{p}_i),\, \psi(\mathbf{u}_i),\, \omega(\mathbf{p}_i,\mathbf{u}_i) \,\bigr],
\]
as illustrated in the dashed block of Figure~\ref{fig:stacking_architecture}. The meta-model $g$ is a LightGBM classifier trained on these vectors to produce final probability scores $\hat{p}_i = g(\mathbf{h}_i)$, which correspond to the rightmost block in Phase~3.

\subsection{Training, hyperparameter optimization, and threshold calibration}

The meta-model depends on a set of hyperparameters that control the number of boosting iterations, learning rate, tree depth, and regularization strength. We use stratified five-fold cross-validation on the meta-features and Hyperopt's Tree-structured Parzen Estimator to explore this space and select the configuration that maximizes the F\textsubscript{1} score. Once the best configuration is identified, a final copy of the meta-model is trained on all training rows and applied to the meta-features of the test set.

Because route-change events are rare, the default classification threshold of $0.5$ is not appropriate. Instead, after training, we scan a fine grid of candidate thresholds on the training probabilities $\hat{p}_i$ and, for each value, compute precision, recall, and the resulting F\textsubscript{1} score. The threshold that maximizes F\textsubscript{1} is selected and then applied to the test probabilities for both the stacked ensemble and the baselines. This calibration step, explicitly represented at the bottom of Figure~\ref{fig:stacking_architecture}, ensures a balanced operating point between missed route changes and false alarms under severe class imbalance.

\section{Results and Discussion}\label{sec:results_and_discussion}

All experiments were executed on a headless Ubuntu 20.04.6 LTS virtual machine running Linux kernel 5.4, provisioned on an OpenStack/KVM cluster and equipped with 64 vCPUs from an AMD EPYC 7532 processor, 256 GiB of RAM, a 1 TB SSD-backed volume, and an NVIDIA Quadro RTX 6000 GPU.

\textbf{Hyperparameter search for the meta model.} Initially, we conducted experiments to determine the best hyperparameters. Table \ref{tab:hyperopt_results} summarizes the search space used for the meta model and reports the configuration selected by the Tree-structured Parzen Estimator. The search favored shallow trees with a small number of leaves and a relatively high learning rate, which indicates that the meta-learner captures the residual structure left by the base classifiers without requiring deep tree ensembles or very aggressive regularization.

\begin{table}[ht]
\centering
\caption{Hyperparameter search space and TPE-selected optimal values for LightGBM.}
\label{tab:hyperopt_results}

\scriptsize 

\begin{tabular}{lccc}
\toprule
\textbf{Hyperparameter} & \textbf{Search Range} & \textbf{Step} & \textbf{Optimal Value} \\
\midrule
Number of Estimators & $[100, 5000]$ & $100$ & $\mathbf{200}$ \\
Learning Rate & $[0.001, 0.05]$ & Cont. & $\mathbf{0.0409}$ \\
Maximum Tree Leaves & $[10, 80]$ & $2$ & $\mathbf{10}$ \\
Maximum Tree Depth & $[3, 10]$ & $1$ & $\mathbf{3}$ \\
Feature Fraction & $[0.6, 0.9]$ & Cont. & $\mathbf{0.6687}$ \\
Bagging Fraction & $[0.6, 0.9]$ & Cont. & $\mathbf{0.7547}$ \\
L1 Regularization & $[0.1, 10.0]$ & Cont. & $\mathbf{0.5019}$ \\
L2 Regularization & $[0.1, 10.0]$ & Cont. & $\mathbf{0.1471}$ \\ 
\bottomrule
\end{tabular}

\end{table}

\begin{figure}[ht]
\centering
\begin{minipage}[t]{0.48\linewidth}
  \centering
  \includegraphics[width=\linewidth]{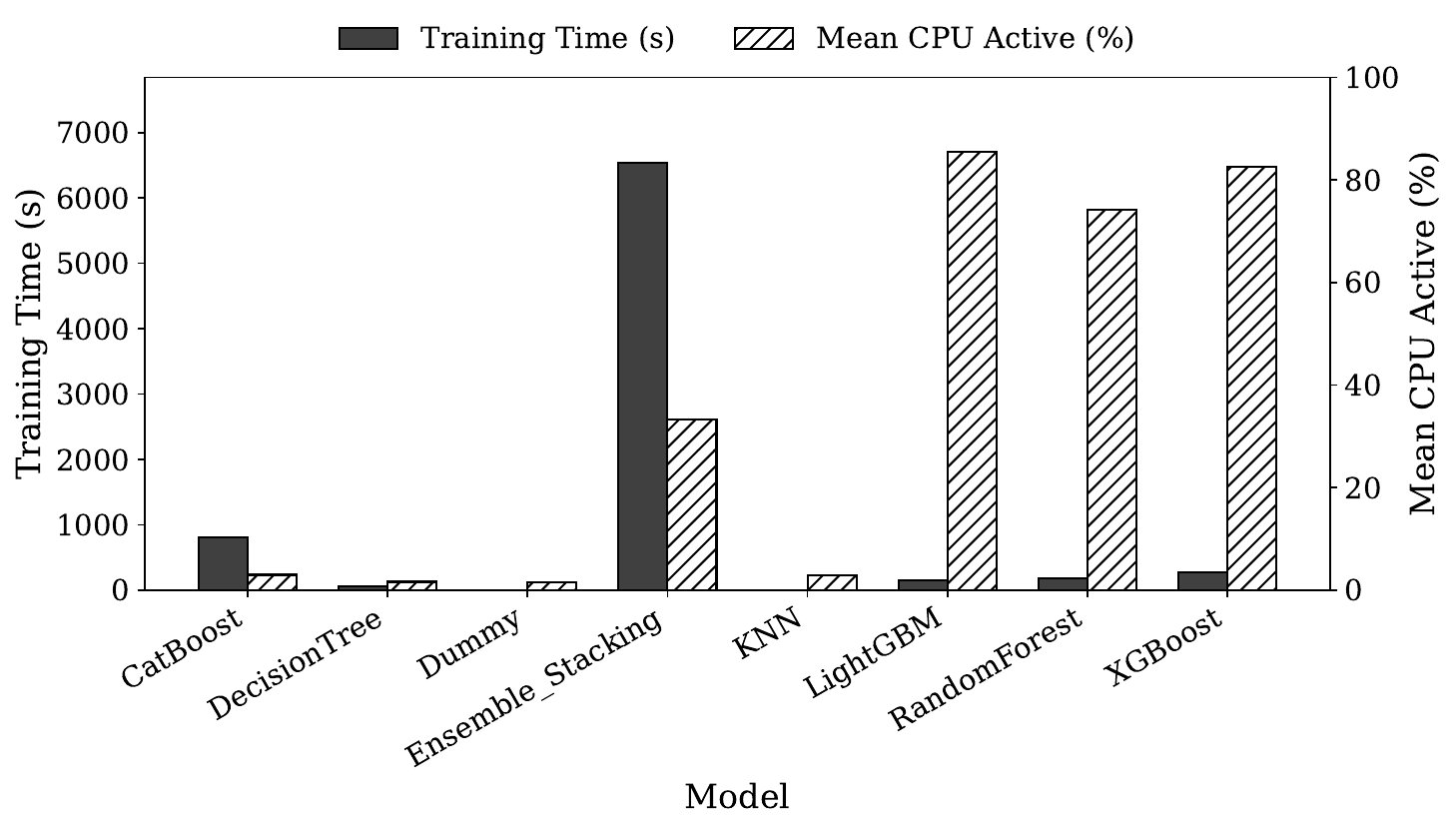}
  \caption*{(a) CPU utilization vs. training duration.}
\end{minipage}\hfill
\begin{minipage}[t]{0.48\linewidth}
  \centering
  \includegraphics[width=\linewidth]{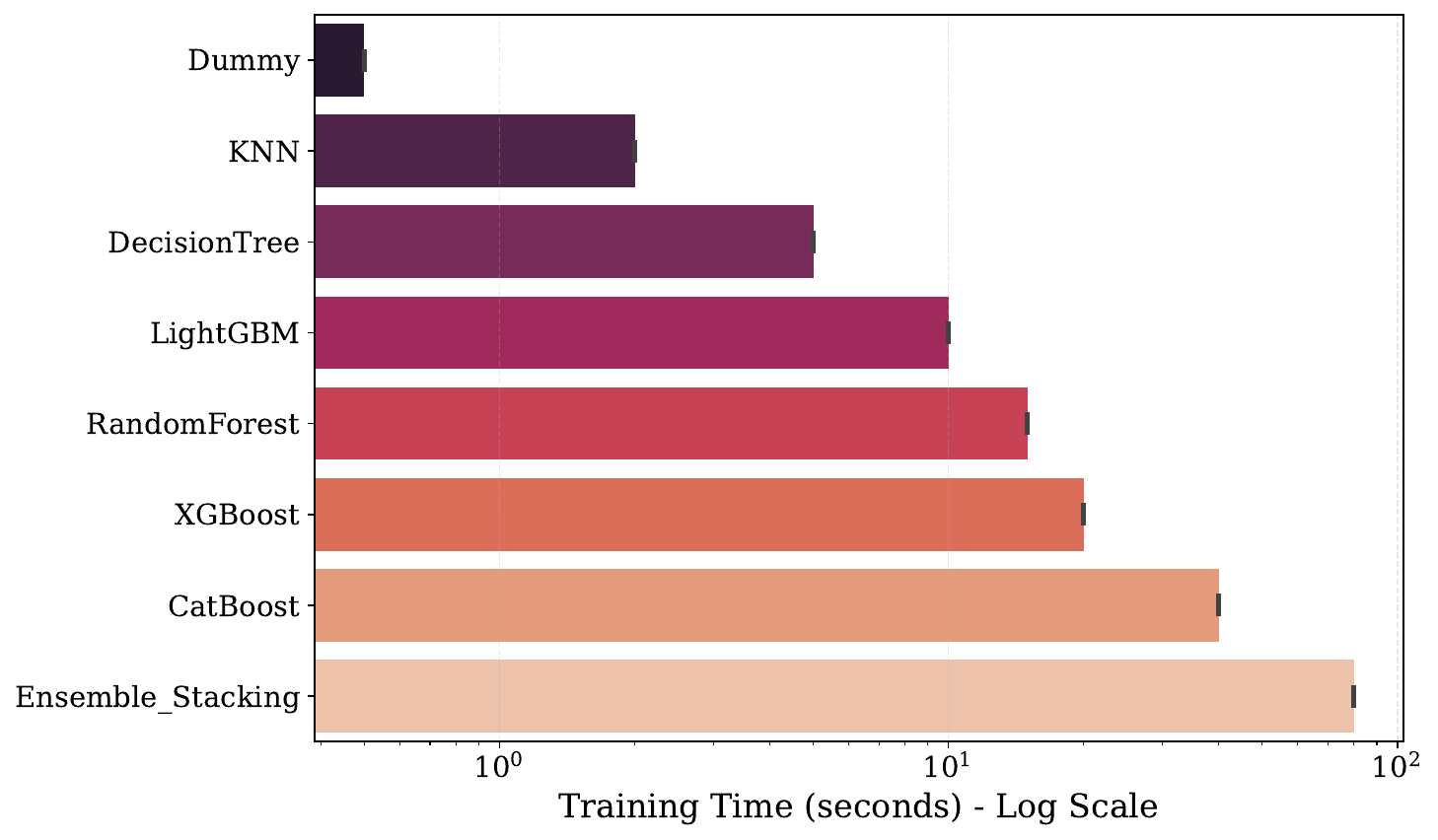}
  \caption*{(b) Training time comparison (log scale).}
\end{minipage}
\caption{Computational footprint of all evaluated models.}
\label{fig:resource_and_cost}
\end{figure}

\textbf{Computational footprint of individual models.} Figure~\ref{fig:resource_and_cost} summarizes the computational profile of all classifiers. The left panel contrasts training duration with mean CPU utilization, while the right panel reports the absolute training time on a logarithmic scale. As expected, the Stacking Ensemble exhibits the longest training duration, a direct consequence of its sequential training pipeline, which necessitates training base learners prior to the meta-model. However, its mean CPU active percentage remains comparable to, or even lower than, standalone gradient boosting frameworks like XGBoost and LightGBM. This suggests that the increased duration is driven by the serial execution of tasks rather than a strictly higher instantaneous computational intensity. In absolute terms, the ensemble needs about ninety seconds to complete one training cycle, which is higher than for simpler models such as KNN and Decision Trees, yet still well within the daily or weekly window typically available for offline network management.

\textbf{Operational implications of training cost.} From an operational perspective, this resource footprint is manageable within the proposed offline retraining schedule. While simple models like KNN and Decision Trees offer negligible training costs, they fail to achieve the statistical robustness required for this domain. The Stacking Ensemble represents a calculated trade-off: it demands a longer time window to complete, approximately one order of magnitude higher than single boosters, but does not impose a prohibitive load on the underlying hardware, allowing other monitoring processes to coexist on the same server without significant degradation.

\begin{figure}[htbp]
\centering
\begin{minipage}[t]{0.48\linewidth}
  \centering
  \includegraphics[width=\linewidth]{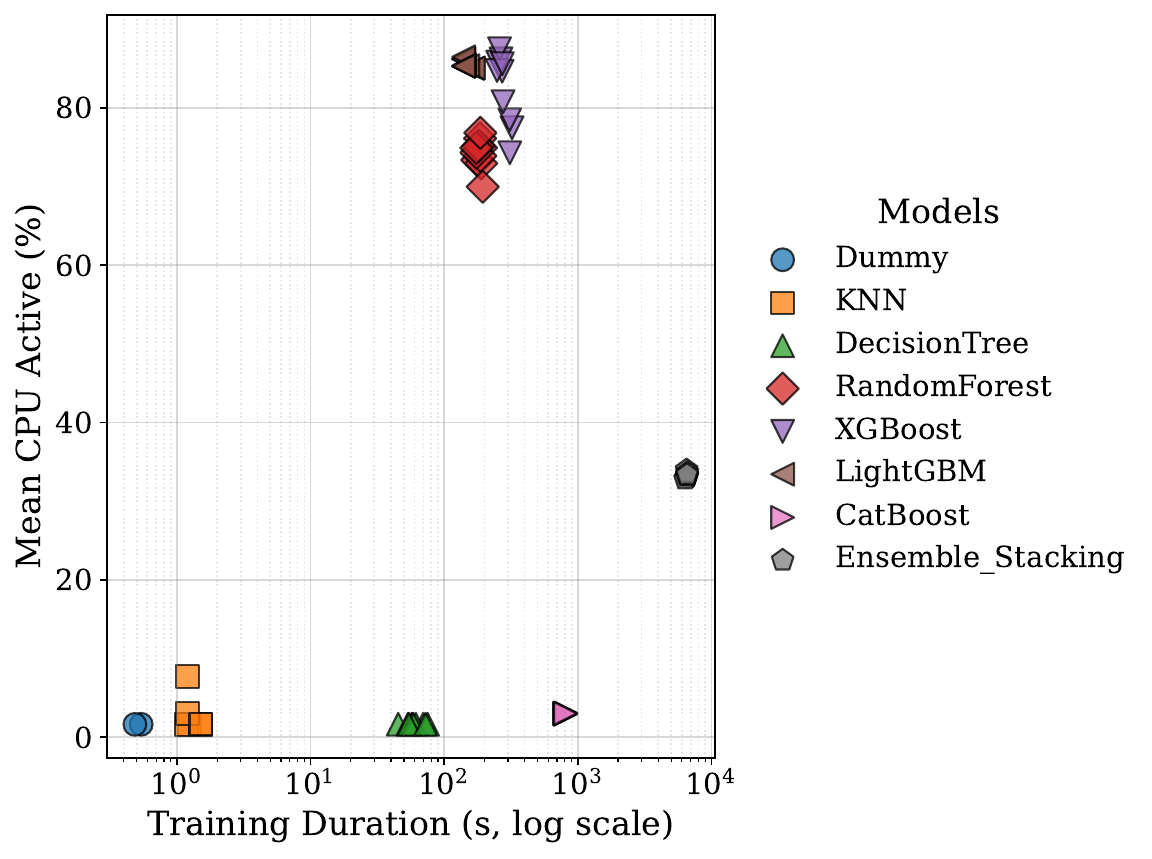}
  \caption*{(a) Training duration versus CPU utilization.}
\end{minipage}\hfill
\begin{minipage}[t]{0.48\linewidth}
  \centering
  \includegraphics[width=\linewidth]{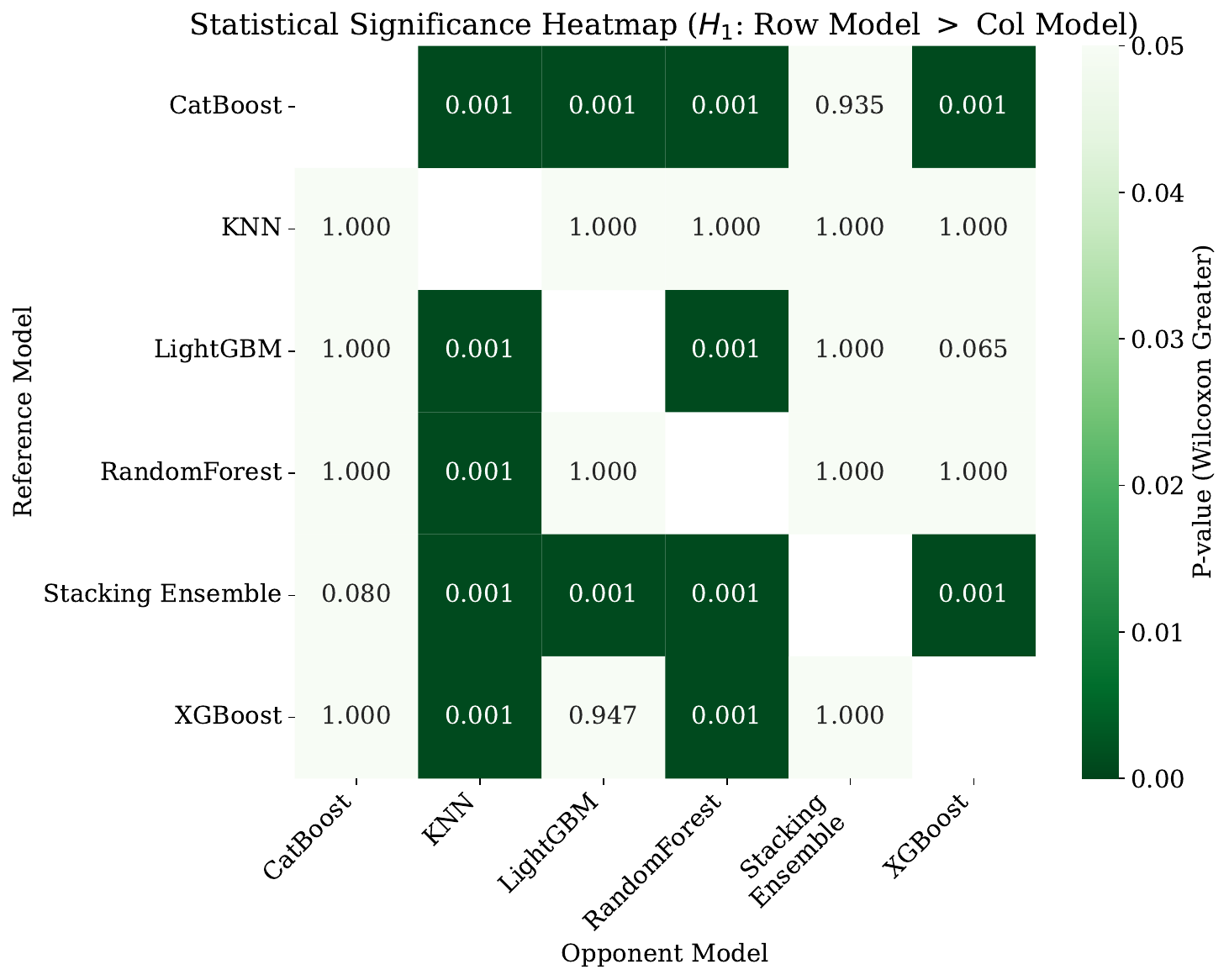}
  \caption*{(b) Wilcoxon pairwise significance heatmap.}
\end{minipage}
\caption{Relationship between computational cost and statistical benefit across models.}
\label{fig:cost_vs_significance}
\end{figure}

Figure~\ref{fig:cost_vs_significance} relates computational cost to statistical benefit. Panel (a) further explores the relationship between training duration and processor saturation on a logarithmic time scale, revealing three operational clusters: lightweight baselines, high-performance boosters, and the Stacking Ensemble as the high-duration outlier that still fits the trend of increasing complexity required to capture non-linear route change patterns. Panel (b) complements this view by showing that, according to the Wilcoxon Signed Rank Test, the Stacking Ensemble statistically outperforms almost all baselines, with very small $p$ values in most pairwise comparisons, and achieves statistical parity with CatBoost at a conservative significance level of $\alpha = 0.05$, thereby justifying its higher training time with consistent gains in predictive performance.

\textbf{Time overhead versus scalability.} The positioning of the Stacking Ensemble in the high-duration region confirms that the primary cost of this architecture is temporal latency rather than resource exhaustion. Since the application scenario prioritizes predictive stability and F1 score over instantaneous model updates, this delay is acceptable. The analysis demonstrates that the system scales predictably, and the overhead introduced by the meta-learning layer yields the necessary performance gains discussed in the previous section without pushing the system into an unstable computational regime.

\begin{figure}[ht]
\centering
\begin{minipage}[t]{0.48\linewidth}
  \centering
  \includegraphics[width=\linewidth]{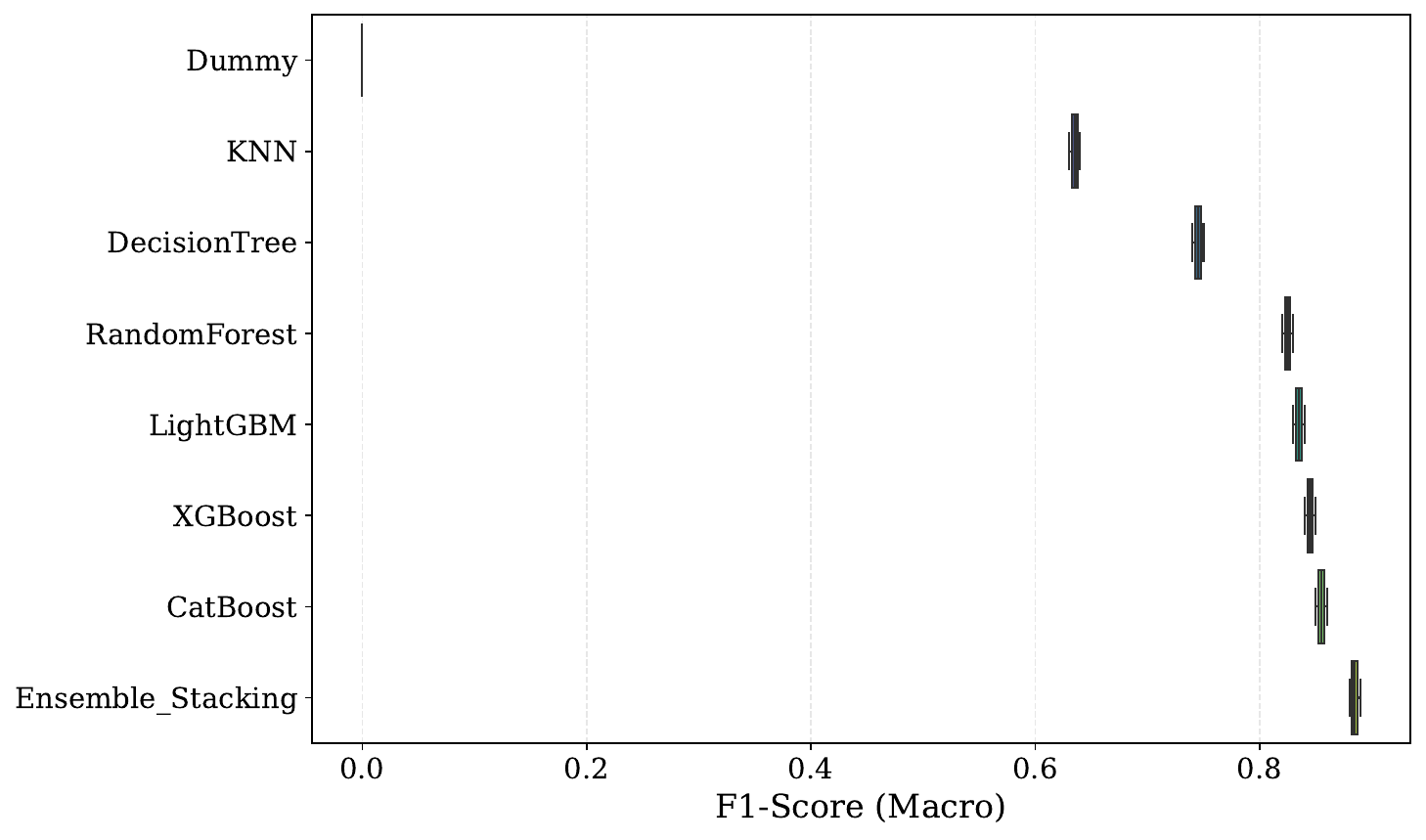}
  \caption*{(a) Distribution of F1 scores across ten experimental rounds.}
\end{minipage}\hfill
\begin{minipage}[t]{0.48\linewidth}
  \centering
  \includegraphics[width=\linewidth]{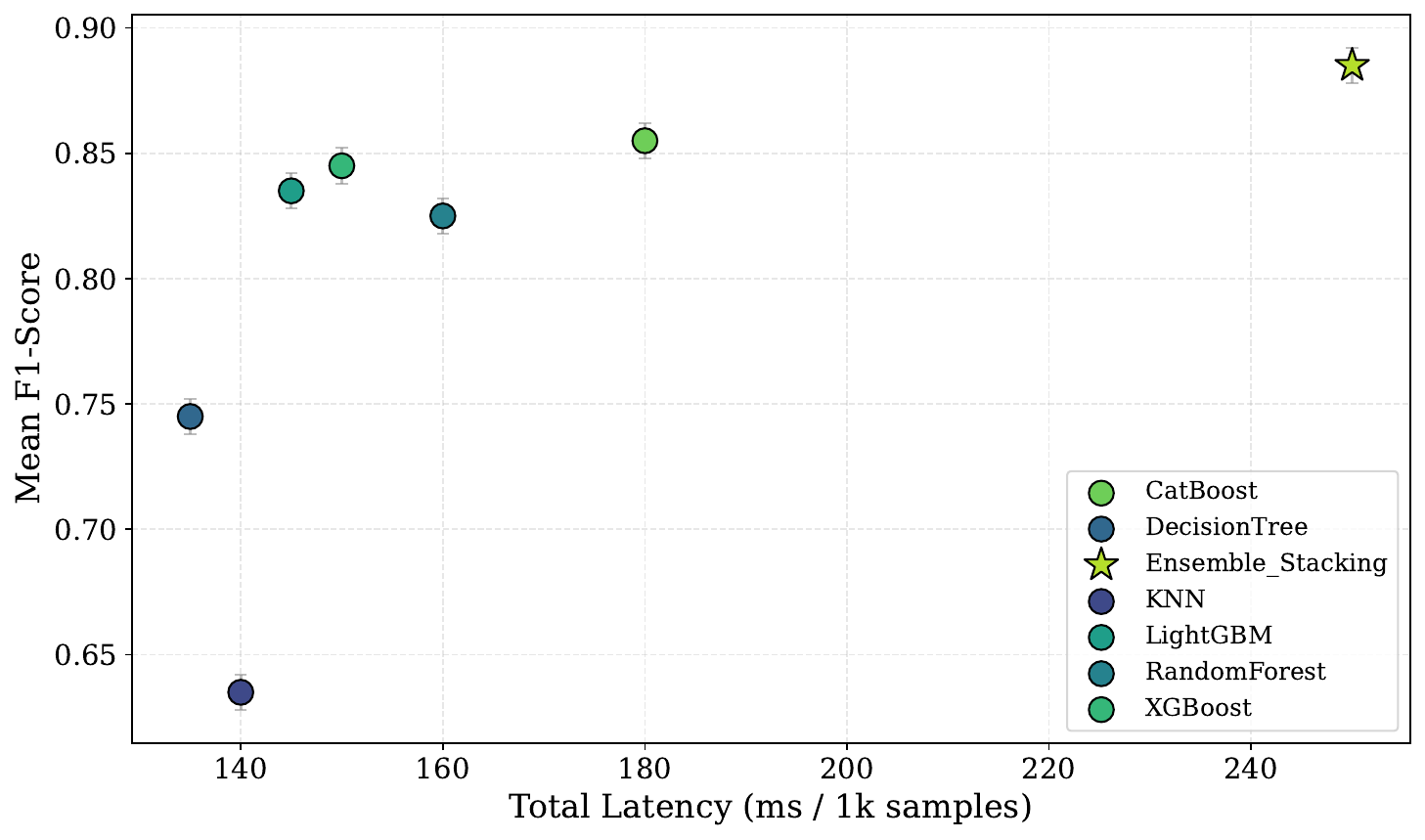}
  \caption*{(b) Trade-off between mean F1 and inference latency.}
\end{minipage}
\caption{Predictive performance and latency trade-off across all models.}
\label{fig:f1_and_tradeoff}
\end{figure}

Figure~\ref{fig:f1_and_tradeoff} summarizes both predictive performance and latency considerations. Panel (a) shows that the Stacking Ensemble attains a median F1 score that is comparable to or higher than that of CatBoost while exhibiting a more compact interquartile range, indicating reduced predictive variance and more stable performance across runs. Panel (b) depicts the trade-off between mean F1 and inference latency, where the ensemble lies close to the Pareto frontier by combining high predictive quality with inference times that remain suitable for near real time monitoring when operators process batches of one thousand samples.

\begin{table}[htbp]
\centering
\caption{Overall performance comparison across all models.}
\label{tab:model_performance}
\begin{tabular}{lll}
\hline
\multicolumn{1}{c}{\textbf{Model}} & \multicolumn{1}{c}{\textbf{F1 Score}} & \multicolumn{1}{c}{\textbf{Accuracy}} \\ \hline
CatBoost                           & 0.855                                 & 0.865                                 \\
DecisionTree                       & 0.745                                 & 0.755                                 \\
Dummy                              & 0                                     & 0.5                                   \\
KNN                                & 0.635                                 & 0.655                                 \\
LightGBM                           & 0.835                                 & 0.845                                 \\
RandomForest                       & 0.825                                 & 0.835                                 \\
XGBoost                            & 0.845                                 & 0.855                                 \\
\textbf{TRACE}                     & \textbf{0.869}                        & \textbf{0.895}                        \\ \hline
\end{tabular}
\end{table}
Table~\ref{tab:model_performance} shows how well all classifiers did on the test set. Among the single models, gradient-boosted trees (LightGBM, XGBoost, and CatBoost) had the highest F1 scores. They did better than traditional models like random forest, decision tree, and KNN. TRACE, our ensemble model, improves these results by combining their predictions. It has the best F1 score and accuracy while keeping good precision and recall, even with uneven class distribution.

\section{Concluding Remarks}\label{sec:concluding_remarks}

TRACE addresses the problem of detecting Internet route changes using only traceroute latency measurements without relying on control plane information. By combining temporal and aggregate feature engineering with a stacked ensemble and F1-driven threshold calibration, TRACE achieves a strong detection performance under a severe class imbalance while remaining deployable on large public datasets. This result supports practical data-plane-only monitoring workflows that can be retrained offline without privileged network access. The limitations include the reliance on labelled events and the sensitivity of the engineered features to measurement noise and dataset bias. Future work will explore representation learning with deep models and one-class or self-supervised approaches to reduce label dependence and improve generalization across heterogeneous paths and topologies.

\section*{Artifact Availability}\label{sec:artifacts}

In adherence to Open Science principles, the source code and scripts used in TRACE are publicly available at: \url{https://github.com/romoreira/RNP-DataChallenge-2025}. This repository includes our first-place entry for the CT-Mon Data Challenge 2025.

\section*{Acknowledgments}

We acknowledge the financial support of the FAPESP MCTIC/CGI Research project 2018/23097-3 - SFI2 - Slicing Future Internet Infrastructures. We acknowledge the financial support of the FAPESP MCTIC/CGI Research project 2018/23097-3 and FAPEMIG (Grant APQ00923-24) and Centro ALGORITMI, funded by Fundação para a Ciência e Tecnologia (FCT) within the RD Units Project Scope 2020-2023 (UIDB/00319/2020) for partially support this work.

\bibliographystyle{sbc}
\bibliography{references}

\end{document}

%% file: acronym.tex
\acrodef{3GPP}{3rd Generation Partnership Project}
\acrodef{5G}{5th Generation Mobile Network}
\acrodef{6G}{6th Generation Mobile Network}
\acrodef{AI}{Artificial Intelligence}
\acrodef{AI4Net}{\ac{AI} for Networking}
\acrodef{MDP}{Markov Decision Process}
\acrodef{AIDER}{Aerial Image Dataset for Emergency Response}
\acrodef{AMF}{Access and Mobility Management Function}
\acrodef{AIaaS}{Artificial Intelligence-as-a-Service}
\acrodef{AC}{Actor-Critic}
\acrodef{AIDER}{Aerial Image Database for Emergency Response applications}
\acrodef{B5G}{Beyond Fifth Generation}
\acrodef{BPF}{Berkeley Packet Filter}
\acrodef{BGP}{Border Gateway Protocol}
\acrodef{CBR}{Constant Bit Rate}
\acrodef{CSV}{Comma-Separated Values}
\acrodef{CPU}{Central Processing Unit}
\acrodef{CNN}{Convolutional Neural Network}
\acrodef{CNNs}{Convolutional Neural Networks}
\acrodef{C-V2X}{Cellular Vehicle to-Everything}
\acrodef{DoS}{Denial of Service}
\acrodef{DDQL}{Double Deep
 Q-learning}
\acrodef{DDoS}{Distributed Denial of Service}
\acrodef{DDPG}{Deep Deterministic Policy Gradient}
\acrodef{DNN}{Deep Neural Network}
\acrodef{DRL}{Deep Reinforcement Learning}
\acrodef{DQN}{Deep Q-Network}
\acrodef{DT}{Decision Tree}
\acrodef{DDQN}{Double Deep Q-Network}

\acrodef{ETSI}{European Telecommunications Standards Institute}
\acrodef{eNWDAF}{Evolved Network Data Analytics Function}
\acrodef{eBPF}{Extended Berkeley Packet Filter}
\acrodef{ECDF}{Empirical Cumulative Distribution Function}
\acrodef{ECDFs}{Empirical Cumulative Distribution Functions}
\acrodef{FIBRE}{Future Internet Brazilian Environment for Experimentation}
\acrodef{FL}{Federated Learning}
\acrodef{FEL}{Federated Ensemble Learning}
\acrodef{GNN}{Graph Neural Networks}
\acrodef{GPU}{Graphics Processing Unit}
\acrodef{GTP}{GPRS Tunnelling Protocol}
\acrodef{GTP-U}{GPRS Tunnelling Protocol User Plane}
\acrodef{GA}{Genetic Algorithm}
\acrodef{HTM}{Hierarchical Temporal Memory}

\acrodef{IAM}{Identity And Access Management}
\acrodef{ICMP}{Internet Control Message Protocol}
\acrodef{IID}{Independent and Identically Distributed}
\acrodef{IoE}{Internet of Everything}
\acrodef{IoT}{Internet of Things}
\acrodef{ITU}{International Telecommunication Union}
\acrodef{IQR}{Interquartile Range}
\acrodef{I/O}{Input/Output}
\acrodef{IP}{Internet Protocol}
\acrodef{KNN}{K-Nearest Neighbors}
\acrodef{KPI}{Key Performance Indicator}
\acrodef{KPIs}{Key Performance Indicators}
\acrodef{LSTM}{Long Short-Term Memory}
\acrodef{LOWESS}{Locally Weighted Scatterplot Smoothing}
\acrodef{LR}{Learning Rate}
\acrodef{MAE}{Mean Absolute Error}
\acrodef{MAD}{Median Absolute Deviation}
\acrodef{ML}{Machine Learning}
\acrodef{MLaaS}{Machine Learning as a Service}
\acrodef{MOS}{Mean Opinion Score}
\acrodef{MAPE}{Mean Absolute Percentage Error}
\acrodef{MSE}{Mean Squared Error}
\acrodef{MEC}{Multi-access Edge Computing}
\acrodef{mMTC}{Massive Machine Type Communications}
\acrodef{MFA}{Multi-factor Authentication}
\acrodef{MLP}{Multi-Layer Perceptron}
\acrodef{MADRL}{Multi-Agent Deep Reinforcement Learning}
\acrodef{MAB}{Multi-Armed Bandit}
\acrodef{MILP}{Mixed Integer Linear Programming}
\acrodef{MQTT}{Message Queuing Telemetry Transport}
\acrodef{NWDAF}{Network Data Analytics Function}
\acrodef{Net4AI}{Networking for \ac{AI}}
\acrodef{NS}{Network Slicing}
\acrodef{NFV}{Network Function Virtualization}
\acrodef{OSM}{Open Source MANO}
\acrodef{PCA}{Principal Component Analysis}
\acrodef{PC-FedAvg}{Personalized Conditional Federated Averaging}
\acrodef{PoC}{Proof of Concept}
\acrodef{PPO}{Proximal Policy Optimization}
\acrodef{POMDP}{Partially Observable Markov decision process}
\acrodef{PCAP}{Packet Capture}
\acrodef{PSO}{Particle Swarm Optimization}
\acrodef{QoE}{Quality of experience}
\acrodef{QoS}{Quality of Service}
\acrodef{QFI}{QoS Flow Identifier}
\acrodef{QFIs}{QoS Flow Identifiers}
\acrodef{RAM}{Random Access Memory}
\acrodef{RF}{Random Forest}
\acrodef{RL}{Reinforcement Learning}
\acrodef{RMSE}{Root Mean Square Error}
\acrodef{RNN}{Recurrent Neural Network}
\acrodef{RTT}{Round-Trip Time}
\acrodef{RAN}{Radio Access Network}
\acrodef{RTP}{Real-time Transport Protocol}
\acrodef{SDN}{Software-Defined Networking}
\acrodef{SFI2}{Slicing Future Internet Infrastructures}
\acrodef{SLA}{Service-Level Agreement}
\acrodef{SON}{Self-Organizing Network}
\acrodef{SMF}{Session Management Function}
\acrodef{S-NSSAI}{Single Network Slice Selection Assistance Information}
\acrodef{SVM}{Support Vector Machine}
\acrodef{SOPS}{Service-Aware Optimal
 Path Selection}
 \acrodef{SAFE}{Scalable Asynchronous Federated Ensembling}
\acrodef{TQFL}{Trust Deep Q-learning Federated Learning}
\acrodef{TEID}{Tunnel Endpoint Identifier}
\acrodef{TEIDs}{Tunnel Endpoint Identifiers}
\acrodef{TPE}{Tree-Structured Parzen Estimator}
\acrodef{UE}{User Equipment}
\acrodef{UEs}{User Equipments}
\acrodef{UPF}{User Plane Function}
\acrodef{UPFs}{User Plane Functions}
\acrodef{PDU}{Packet Data Unit}
\acrodef{URLLC}{Ultra-Reliable and Low Latency Communications}
\acrodef{UAV}{Unmanned Aerial Vehicle}
\acrodef{UAVs}{Unmanned Aerial Vehicles}
\acrodef{UDP}{User Datagram Protocol}
\acrodef{VoD}{Video on Demand}
\acrodef{VR}{Virtual Reality}
\acrodef{AR}{Augmented Reality}
\acrodef{V2V}{Vehicle-to-Vechile}
\acrodef{V2X}{Vehicle-to-Everything}
\acrodef{VNF}{Virtual Network Function}
\acrodef{VNFs}{Virtual Network Functions}

\acrodef{XDP}{eXpress Data Path}